\documentclass{aa}

\include{epsf}

\begin{document}

   \thesaurus{03 (13.18.1; 11.01.2; 11.03.2; 11.05.2; 11.09.1 2021+614)}

   \title{The GHz-Peaked Spectrum radio galaxy \object{2021+614}:
          Detection of slow motion in a compact symmetric object}

   \titlerunning{2021+614: Detection of slow motion in a CSO}

   \author{W. Tschager \inst{1}
   \and    R.T. Schilizzi \inst{2}$^,$\inst{1}
   \and    H.J.A. R\"ottgering \inst{1}
   \and    I.A.G. Snellen \inst{3}
   \and    G.K. Miley \inst{1}}

   \offprints{W. Tschager\\ \emph{email:} tschager@strw.LeidenUniv.nl}

   \institute{Leiden Observatory, P.O. Box 9513, 2300 RA  Leiden, The
              Netherlands
   \and       J.I.V.E., P.O. Box 2, 7990 AA Dwingeloo, The Netherlands
   \and       Institute of Astronomy, Madingley Rd, Cambridge CB3~OHA, UK}
    
   \date{Received April 18, 2000; accepted June 28, 2000}

   \maketitle

\begin{abstract}

We have analysed VSOP (VLBI Space Observatory Programme) data at 5~GHz and
ground-based VLBI (Very Long Baseline Interferometry) data at 15~GHz for
the GHz-Peaked Spectrum (GPS) radio galaxy 2021+614. Its morphology is
consistent with it being a compact symmetric source extending over
$30\,h^{-1}$ pc. From a comparison with earlier observations we have
detected an increase in the separation and a decrease in the size of the
two most prominent components. We determine the projected speed with which
these two components recede from each other to be $0.12 \pm
0.02\,h^{-1}\,c$. Given the projected separation of the two components of
$16.1\,h^{-1}$ pc, the infered kinematic age is $440 \pm 80$ years,
measured in the source reference frame\footnote{For all calculations
involving cosmological models we use $H_0=100\,h\,{\rm
km\,s^{-1}\,Mpc^{-1}}$, $q_0 = 0.5$.}. These results provide additional
support for the contention that compact symmetric radio objects are young
and the precursors of the classical FR~I or FR~II radio sources. The sizes
of individual components appear to contract with time which is not
consistent with the self-similar evolution model for peaked spectrum
sources.\\
In order to overcome problems related to the estimation of uncertainties
for separation measurements between source components, we have developed
and applied a method that compares two uv-data sets obtained at different
epochs. This method parametrizes the most important structural change, the
increase in separation between components, by rescaling the u and v axis
of the amplitude interference pattern. It provides best-fit values for the
parameters and uses a bootstrap method to estimate the errors in the
parameters.


\end{abstract}

\section{Introduction}

\begin{figure*} 
\epsfxsize=17.9cm \epsfbox{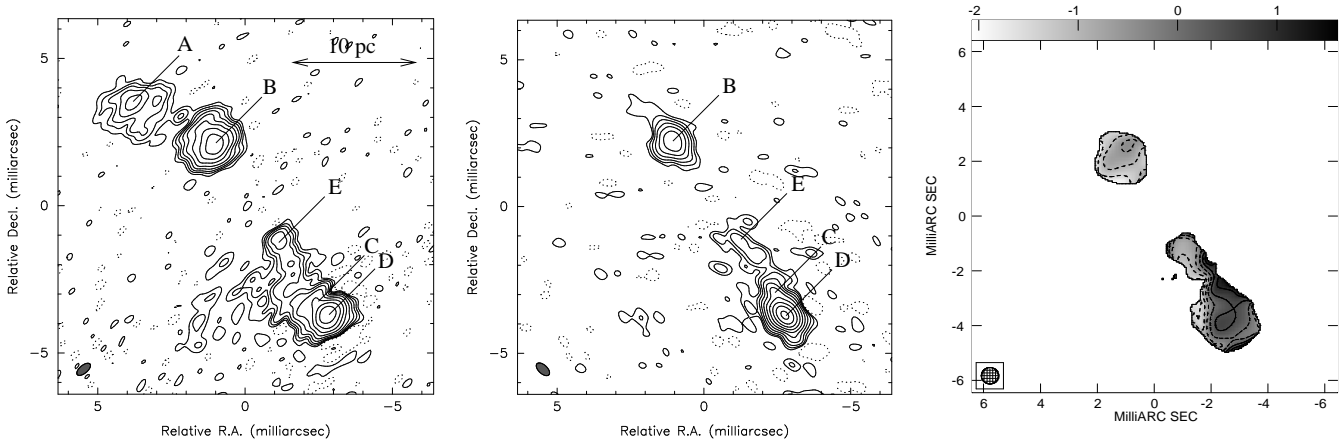}
\caption[]{5 GHz VSOP image (left) and 15 GHz global VLBI image (middle).  
The rms noise is 0.5 mJy/beam for the 5 GHz image and 0.9 mJy/beam for the
15 GHz image. Image peaks are 0.475 Jy/beam and 1.12 Jy/beam,
respectively.  Contour levels are drawn at -2, 2, 4, 8, 16, \dots, 512,
(1024) $\times$ the respective rms noise.  The restoring beams are $0.56
\times 0.27$ mas, $-50.3\degr$ and $0.56 \times 0.30$ mas, $47.4\degr$,
respectively. At $z = 0.2266$, 1 mas corresponds to $2.30\,h^{-1}$ pc.  
(Right) Image of the 5 -- 15 GHz spectral-index, $\alpha$ ($S_\nu \propto
\nu^{\alpha}$). The spectral index was calculated in regions with
intensity higher than 5-$\sigma$ in both images.  Contour levels are drawn
at spectral index values of -1.5, -1.0, -0.5, 0.0, 0.5 and 1.0; negative
levels and the zero level are dashed. The restoring beam is 0.6 mas
circular.}
\label{fig1} 
\end{figure*}

Despite many years of study of extragalactic radio sources, it is still
unclear how they are formed and evolve. A crucial element in the study of
their early evolutionary stages is to identify the young counterparts of
old and extended FR~I/FR~II objects. Good candidates for young radio
sources are those with peaked spectra, GHz-Peaked Spectrum (GPS) sources
and Compact Steep-Spectrum (CSS) sources, because they are small in
angular size as expected for young sources. GPS sources are characterized
by a simple convex radio spectrum peaking at a frequency of about 1 GHz
and are typically 100 pc in size. CSS sources have peaks in their spectra
at lower frequencies and have projected linear sizes of $<15$ kpc.\\
The best direct evidence for very low kinematic ages has now been found
for a few GPS radio galaxies. Measurements of hotspot advance speeds, from
which an age estimate can be made, have been obtained for 0108+388
(Owsianik et al. \cite{Owsianik2}), 0710+439, 2352+495 (Owsianik \& Conway
\cite{Owsianik1}; Owsianik et al. \cite{Owsianik3}) and 1943+456
(Polatidis \cite{Polatidis}). The hotspot advance speeds are typically of
order $0.1\,h^{-1}\,c$, translating into ages of a few hundred to a few
thousand years.  All these sources belong to the morphological class of
Compact Symmetric Objects (CSO), which are characterized by their small
size ($< 500$ pc) and symmetric radio morphology.\\
Having established that CSOs are young it is interesting to examine them
at a number of points along their evolutionary track in the $P$--$D$
(luminosity -- linear size) diagram, and to carry out investigations at
several radio wavelengths with a range of angular resolution and limiting
flux density.  We are investigating a sample of 11 bright GPS sources at 5
GHz with VSOP.  These observations have been complemented by VLBI
observations at 15 GHz to obtain matched-beam spectral index data. The 11
GPS sources in our sample are all those known in November 1995 with
declination $\delta > 25\degr$, peak frequency $\nu_{\rm peak} \sim
5$~GHz, and peak flux-density $S_{\rm peak}$ $ ^>_\sim$ 0.5~Jy/beam. \\
Here we report on one such object, the GPS radio galaxy 2021+614, also
named OW 637. It is one of the strongest GPS sources with a total flux
density of 2.5 Jy at 5 GHz and a radio luminosity of $6 \times
10^{37}\,h^{-2}$ W between 100 MHz and 100 GHz.  The radio spectrum of
2021+614 has a broad, relatively flat peak centred at about 4 GHz, and
falls off at lower and higher frequencies. The flattening of the spectrum
at the highest radio frequencies ($\sim 100$ GHz) indicates the presence
of a very compact component (Steppe et al. \cite{Steppe}).  The flux
density above the spectral peak shows variability. Seielstad et al.  
(\cite{Seielstad}) detect a 20\% total flux density change on a time scale
of 10 years. Aller et al. (\cite{Aller}) observe some variability at 15
GHz, but not at 5 GHz, on a time scale of 5 years.  High
angular-resolution VLBI observations of 2021+614 at 2.3, 5 and 8.4~GHz
have been published by Wittels et al. (\cite{Wittels}), Bartel et al.
(\cite{Bartel2}), Pearson \& Readhead (\cite{Pearson}) and Conway et al.
(\cite{Conway}). Based on the radio morphology and decomposition of the
radio spectrum into contributes from individual components these authors
all prefer a core-jet classification for 2021+614. In addition, Conway et
al.  (\cite{Conway}) investigated structural changes in the source and
identified the apparent centroid shift of the two main components as real
motion. Cawthorne et al. (\cite{Cawthorne}) determined that there is no
significant linearly polarised emission from any of the components at
5~GHz, with upper limits of 5 mJy. The source was also observed by
Kellermann et al. (\cite{Kellermann}) as part of a 2-cm VLBI survey.\\
The optical counterpart of 2021+614 is an elliptical galaxy at redshift
0.2266. It is a highly reddened Narrow Line Radio Galaxy (NLRG) most
probably with a considerable dust component within the optical object
(Bartel et al. \cite{Bartel1}). The shape of the [OIII] ($\lambda_0=5007
$\,\AA) emission line profile is asymmetric and has a velocity dispersion
of 780 km/sec. Deep CCD imaging by O'Dea et al. (\cite{O'Dea1}) shows that
the galaxy has a prominent compact nucleus and two possible companions
within $12\arcsec$.\\
In this paper we report new data from VSOP and global VLBI observations.
Combining these data with older VLBI observations we determine the
increase in separation between the two strongest components first detected
by Conway et al. (\cite{Conway}). In Sect. 2 we describe our observations
and the data used to quantify this increase. In Sect. 3 we present the
morphology and spectral-index distribution for 2021+614, and we calculate
the separation rate of the components. We introduce a complementary
approach for studies of changes in source structure, transfering the
problem from the image coordinate plane into the spatial frequency plane.
Further discussions of the results are in Sect. 4 where we propose that
the morphological classification for the source is indeed CSO rather than
core-jet and deduce its age. In Sect. 5 we summarize our main conclusions.
In Appendix A we elaborate on the problems encountered in deducing the
separation rate and its uncertainty and develop a method which allows to
measure the relative increase in separation occuring between two epochs by
means of the amplitude interference patterns.

\section{Observations and Data Reduction}

The VSOP satellite HALCA observed 2021+614 at 5~GHz on November 6, 1997
together with a 15-station ground-based array composed of all 10 VLBA and
5 of the 10 EVN radio telescopes (Effelsberg, Medicina, Noto, Onsala and
Torun) plus the VLA in phased-array mode. The on-source time was 9 hours
for the ground telescopes and 6 hours for the satellite. The tracking
stations used to downlink the data and relay the local oscillator signal
to the satellite were located at the Deep Space Network sites at Goldstone
in California (USA) and Tidbinbilla in Australia. The VLBI observing run
at 15~GHz took place on October 9, 1998 using the VLBA and the 100-m
Effelsberg radio telescope for three 30-minute scans over a range of hour
angles. Both data sets were correlated at the NRAO Array Operations Center
in Socorro, NM, USA.\\
The 5 and 15 GHz VLBI images are shown in Fig. \ref{fig1} (left and
middle).  The VSOP image was obtained following standard procedures for
editing, \emph{a-priori} amplitude calibration and fringe-fitting as
recommended by the AIPS Cookbook (NRAO AIPS package) for space-VLBI data.  
Imaging was carried out with the Caltech Difference Mapping program
(Difmap, Shepherd et al. \cite{Shepherd}) applying uniform weighting to
the data points. In order to produce a dynamic-range limited image the
data had to be taken through several iterations of phase, and phase and
amplitude self-calibration.  Because of their low sensitivity, baselines
between HALCA and the small ground telescopes needed extensive flagging
during imaging. Relatively high-SNR fringes on space baselines can be only
seen between HALCA and the 100-m Effelsberg radio telescope, and the
phased VLA which has the equivalent sensitivity of a single 115-m antenna.
The longest baseline measures 524 M$\lambda$ corresponding to a resolution
of 0.3 mas (uniform weighting) and was achieved between the VLBA antenna
located at Saint Croix in the Virgin Islands and HALCA.

\begin{figure*}
\epsfxsize=17.9cm \epsfbox{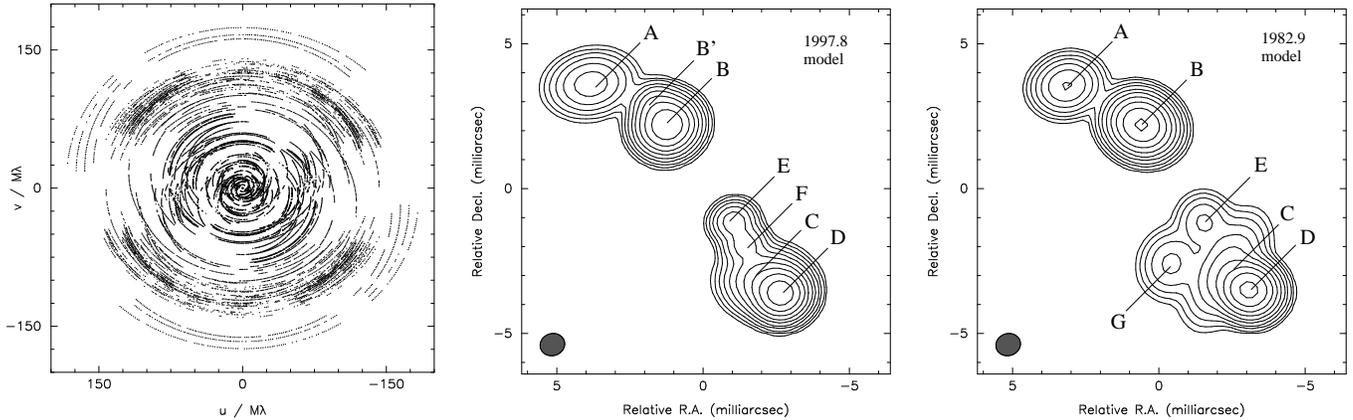}
\caption{The uv-coverage (left) not including space baselines and the
seven-component model (middle) for the 1997.8 observation at 5 GHz.
(Right) Conway's six-component model for the 1982.9 data at 5 GHz.
Contours are drawn at 1, 2, 4, \dots, 512 $\times$ 1 mJy. The convolving
beam is 0.85 $\times$ 0.76 mas at position angle $-70.2\degr$ for both
model images.}
\label{fig2}
\end{figure*}

\section{Results}

\subsection{The sub-mas morphology}

From Fig. \ref{fig1} we see that 2021+614 has a simple symmetric structure
at 5 and 15~GHz, dominated by two bright components.  These two components
are labeled B and D, following Bartel et al. (\cite{Bartel2}). A third
component, labeled A, is visible only on the 5 GHz image, indicating that
it has a steep spectral index ($\alpha < -2$). In addition, a central
component, E, is visible in both images and a jet-like feature connects
this central component to components C and D.  The jet-like feature
appears bent in the 5 GHz image, but not in the 15 GHz image. Low surface
brightness, extended structure can be seen east of component C and D. All
components appear to be resolved to some degree, except the central
feature.  Low level side lobes near component D are due to the sparse
sampling of the uv plane in region between Earth-Earth and Earth-HALCA
baselines and limit the dynamic range of the image.

\subsection{The 5 GHz -- 15 GHz spectral-index distribution}

The distribution of the spectral index $\alpha$ ($S_\nu \propto
\nu^{\alpha}$) shown in Fig. \ref{fig1} (right) is derived from the 5 and
15~GHz images.  Both images were restored with a 0.6-mas circular beam and
the spectral index calculated within regions whose intensity was higher
than $5$ $\times$ rms noise in both images. Three regions with different
spectral index characteristics can be seen: a steep spectrum ($\alpha \sim
-0.7$) north-eastern component, an inverted spectrum ($\alpha \sim 0.4$)  
south-western complex and a central component with steep spectral index
($\alpha \sim -0.6$). The inverted-spectrum border around the
south-western complex is most probably artificial and due to the
differences in beam position angle of the two combined images and possible
noise effects in these low intensity areas.

\subsection{Modelfitting}

\begin{table}
\caption[]{Best-fit model parameters for the 1997.8 observation at 5 GHz.
Component parameters are integrated intensity ($I$), polar coordinates for
component position ($r$, $\Theta$) , major axis FWHM ($b$), axial ratio
($b/a$) and position angle ($\phi$) measured from North through East. The
type $T$ codes ``d'' and ``g'' refer to delta and gaussian components,
respectively.}
\begin{tabular}{ l c c c c c c r}
\hline
      & $I$   & $r$ & $\Theta$ & $b$  & $b/a$ & $\phi$  & $T$ \\
      & Jy    & mas & $\degr$  & mas  &       & $\degr$ &     \\ 
\hline
A     & 0.123 & 9.66 & +41.7   & 1.24 &  0.64 & $-80.8$ & g   \\
B     & 0.841 & 6.94 & +33.5   & 0.68 &  0.98 & $-18.9$ & g   \\
B'    & 0.054 & 7.80 & +33.1   &      &       &         & d   \\
C     & 0.360 & 0.55 & +49.4   & 1.16 &  0.73 &  +80.3  & g   \\
D     & 1.078 & 0.08 &$-117.8$ & 0.50 &  0.87 &  +32.8  & g   \\
E     & 0.053 & 2.92 & +32.2   &      &       &         & d   \\
F     & 0.049 & 2.03 & +35.2   &      &       &         & d   \\
\hline
\label{tab1}
\end{tabular}
\end{table}

As mentioned above we observed 2021+614 at 5 GHz with VSOP and at 15 GHz
with the VLBA and Effelsberg; additionally we had three 15-GHz uv-data
sets from snapshot observations made by K.I. Kellermann et al.
(\cite{Kellermann}). We also make extensive use of Conway's six-component
model for 5 GHz data taken in November 1982 (priv. comm., Conway et al.
\cite{Conway}).\\
Before modelfitting can be performed, the weights of the single visibility
measurements must be determined. This guarantees a meaningful significance
for the reduced $\chi^2$ as an indicator for the goodness of fit. For each
integration time the weight is the reciprocal amplitude variance for the
visibility measurement and is obtained from the internal scatter of the
data within the averaging interval. The averaging time is limited by the
coherence time, which for earth-based VLBI observations at 5 and 15 GHz is
well above the 2-min averaging time adopted. In addition, to avoid
complicated models with numerous components for the 5 GHz VSOP
observation, the highest spatial frequencies were excluded.\\
We performed modelfitting using the Difmap \verb+modelfit+ program. This
program fits the Fourier-transformed image-plane model to the real and
imaginary parts of the complex visibilities (in contrast to other programs
which fit the visibility amplitudes and closure phases) and tacitly
assumes that the visibility phases are well calibrated. Thus, we
modelfitted uv-data sets which have been self-calibrated beforehand.\\
The modelfitting provides a description of the most prominent
characteristics of the brightness distribution -- the number of
components, their position, size, shape and intensity of source components
-- with as few parameters as possible.\\
For each uv-data set we followed the same modelfitting procedure. The
starting model contained two circular gaussians each with a
full-width-at-half-maximum (FWHM) of 0.7 mas located at the position of
the highest pixel of the component. After the initial cycle of
modelfitting additional components were added to improve the fit.
Parameters of all components were allowed to vary during the process.  
This procedure was repeated until the reduced $\chi^2$ did not decrease
any further.  In order to keep the number of parameters needed to fit the
brightness distribution as small as possible we used elliptical gaussian
components as well as components represented by delta functions.\\
For a model that is a good approximation to the data, the expected value
of reduced $\chi^2$ should be about 1. The reduced $\chi^2$ characterizing
our best fitting models (e.g. for the 5 GHz observation) was never less
than 2.7. This indicates that the fit is poor. The reason for this is most
likely that our simple models of a small number of components do not
reproduce the extended, low surface brightness emission. However, this is
not important in investigating the intensity and motion of the main
components.\\
One of the problems in modelfitting is that more than one minumum in the
$\chi^2$ function may exist. We investigated the parameter space for
pathological behaviour of the $\chi^2$ function around the fitted global
minimum, as a function of the most important parameters. Nearby local
minima occur when the intensity or shape of one of the components
degenerates. The associated source models can be rejected on the basis of
their containing negative or unnaturally elongated components. The minima
found for all uv data sets could be identified as being global.\\
For the 15 GHz observations the modelfitting procedure yielded models with
five components, whereas for the 5 GHz observation a seven-component model
best met the requirements. The best fitting model parameters for the 5 GHz
data set are listed in Table \ref{tab1}. Fig. \ref{fig2} (left) shows
visibilities for Earth baselines sampled during the 1997.8 observation at
5 GHz and used for modelfitting. The contour image shown on the right of
the uv-coverage represents the sum of the seven components in the model.
The rightmost panel in Fig. \ref{fig2} shows Conway's six-component model
for the 5 GHz observation from 1982.9.

\subsection{Separation rate}

\begin{figure}
\hbox{\epsfxsize=8.7cm \epsfbox{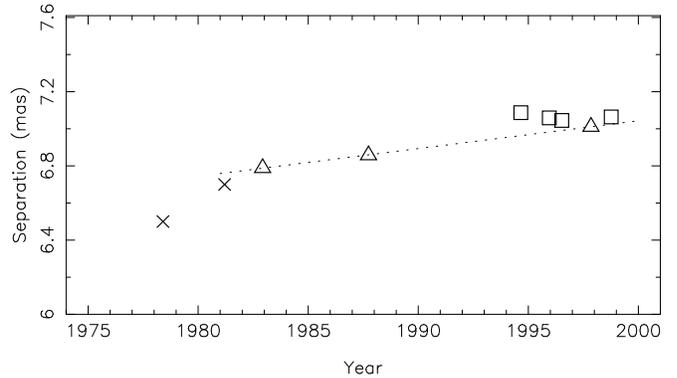}}
\caption{Separation between B and D, the two brightest components, at 5
GHz. Triangles, crosses and squares indicate data from observations at
5~GHz, 8.3~GHz and 15~GHz, respectively. Component separations at 1978.4
and 1981.2 are from Wittels et al. (\cite{Wittels}) and Bartel et al.  
(\cite{Bartel2}), respectively. The 1982.9 and 1987.7 data points are from
Conway et al. (\cite{Conway}). We obtained new separation measurements at
epoch 1994.6, 1996.0, 1996.5 (uv-data from R.C. Vermeulen and K.I.  
Kellermann), 1997.8 (VSOP observation) and 1998.8 (our 15 GHz
observation).  The dotted line is the best fit to the 5 GHz data points
and its slope indicates a separation rate of $14.9 \pm 0.2$ $ \mu$as/yr.
The error reported here represents the scatter of the points about the
best fit line.}
\label{fig3}
\end{figure}

We are interested in measuring the separation between the two brightest
source components and changes of that separation in time. Fig. \ref{fig3}
incorporates all the separation measurements between components B and D,
the two brightest components at 5 GHz, available to us as a function of
observing year.  The linear regression fit to the 5 GHz data points
(triangles) shows that the component separation is changing at a rate of
$14.9 \pm 0.2$ $\mu$as/yr. In the introduction to Appendix A we discuss in
more detail the estimation of the uncertainty for this measurement.\\
The four separation measurements at 15 GHz (squares)  do not show the same
progressive increase of separation seen in the 5 GHz data, whereas the 8.3
GHz data points (crosses) are consistent with separating components.\\
Striking evidence that an increase in separation has occurred in 2021+614
can be seen in Fig. \ref{fig4} where we plot the visibility amplitudes
versus projected uv-distance parallel to the axis defined by the
components B and D. The double source structure along this line can be
seen clearly, as can the effect of the extended nature of the individual
components. Fig. \ref{fig4} (top) shows the self-calibrated visibility
amplitudes from our VSOP observation at epoch 1997.8 out to a projected
baseline length of 180 M$\lambda$. Fig. \ref{fig4} (middle) shows
projected model visibility amplitudes for the data shown in the upper
panel. Fig. \ref{fig4} (bottom) represents Conway's best model for the
1982.9 observation at uv-loci identical to those sampled during the 1997.8
observation. Comparing the fringe patterns from the upper and middle panel
we recognize the missing flux density at short baselines resulting from
unmodelled extended structure - an issue we discussed shortly in section
3.3.\\
Minima in the uv-plane occur at spatial frequencies of $u_{\mathrm{min},n}
= (2n-1)/(2d)$ for the $n^{\rm th}$ minimum measured along the position
angle of the line connecting the two components, where $d$ is the distance
between the two components in radians (Fomalont \& Wright
\cite{Fomalont}). For the 1997.8 data this implies that the distance from
the origin of the uv-plane to the first minimum is $14.7$ M$\lambda$,
whereas for the 1982.9 observation the position of the first minimum
occurs at $15.2$ M$\lambda$. The effect of the increase in separation is
more easily recognizable for the higher order minima: it is qualitatively
evident that an inward shift in the position of the minima has occurred,
as expected for a separating source.\\
The fact that a small increase in separation between source components
translates into an easily measurable change in the interference pattern
led us to develop a method which compares two uv-data sets obtained at
different epochs directly, parametrizing the time evolution, i.e., the
structural change of the source. This method helps to overcome the
problems connected with the estimation of errors for distance and
separation-rate measurements.  These problems are outlined in section
\emph{A.1}.  The critical points of our method are the feasibility and
realization of the direct comparison of the uv-data. In the sections
following \emph{A.1} we describe this procedure and apply it to our data.
The method calculates the two-dimensional factorized increase in
separation directly in the uv-plane and provides error estimates for those
numbers.\\
We find that the visibility-amplitude interference pattern for the 1997.8
observation has to be stretched by $4.43 \pm 0.05$\% in the u-direction
and by $2.12 \pm 0.10$\% in the v-direction in order to overlap with the
1982.9 observation.\\
The fact that the two multiplicative factors differ from each other tells
us that the separation is not a simple linear increase along the line
defined by the position of the two components at epoch 1982.9. The angular
polar coordinate $\psi_2$ of component B with respect to D at epoch 1997.8
is related to that at epoch 1982.9, $\psi_1$, by
$\tan\psi_2=s_u/s_v\,\cdot\,\tan\psi_1$; $s_u$ and $s_v$ are the
stretching factors determined in the appendix. Note that the angular polar
coordinate of component B changes from $32.9\degr$ to $33.5\degr$ between
the two epochs (see Fig. \ref{fig4}).

\subsection{Variability}

\begin{figure}
\hbox{\epsfxsize=8.7cm \epsfbox{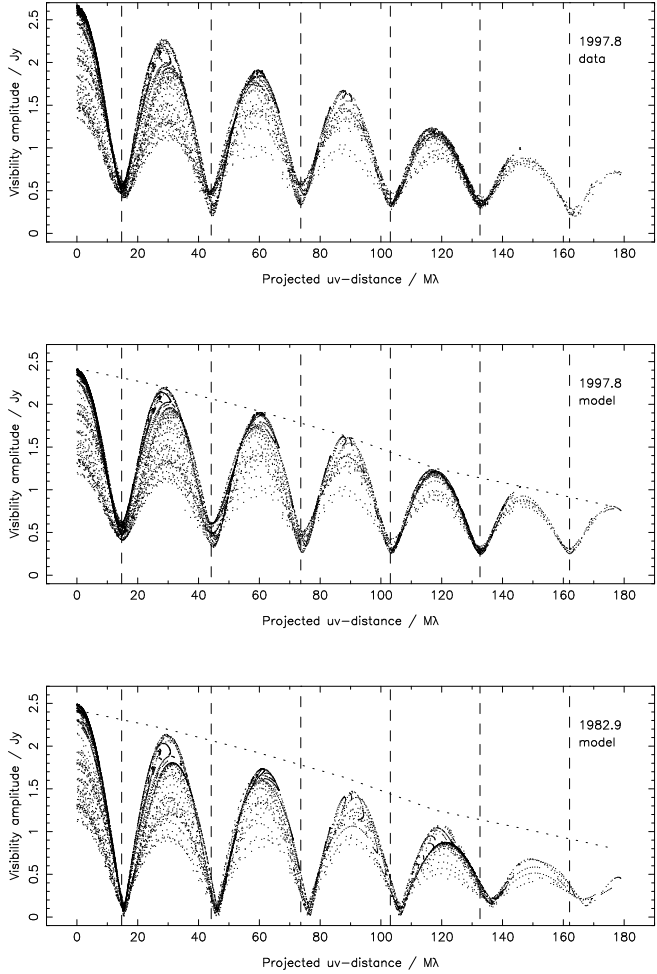}}
\caption{ The upper panel shows self-calibrated visibility amplitudes for
epoch 1997.8 data vs projected uv-distance along position angle
(\emph{p.a.})  33.5$\degr$. The middle and lower panel show projected
model visibility amplitudes from our model (fig. \ref{fig4}, middle) along
\emph{p.a.} 33.5$\degr$ and Conway's model (fig. \ref{fig4}, right) along
\emph{p.a.} 32.9$\degr$, respectively. In all three panels the vertical
lines (dashed) at odd multiples of 14.7 M$\lambda$ mark the positions of
minima as expected from modelfitting to epoch-1997.8 data. The dotted line
in the lower panels marks the upper envelope of the fringes at epoch
1997.8 and is a function of the sizes and shapes of the components.}
\label{fig4} 
\end{figure}

Additional qualitative differences in source characteristics between the
1987.9 and 1997.8 observations can be directly established from Fig.  
\ref{fig4} and quantified using the component models. Extrapolating the
measured visibility amplitudes at low uv-spacing down to zero uv-spacings
gives the total flux density of the source.  There appears to be a
decrease in intensity of about 4\% over the 15 yr period. However, this
may be due to amplitude calibration errors which can be as high as 5\%.\\
On the other hand, it is immediately evident from Fig. \ref{fig4} (middle
and lower panel) that a change in relative intensity of the two brightest
components has occurred. In Conway's model for the 1982.9 observation the
ordinate value of the minima are close to zero, indicating two major
components with almost equal intensity, beating against each other.
Fifteen years later the components are no longer equal in intensity and
consequently the minima lie well above zero.  The comparison of the source
models provides an explanation in terms of changes in component intensity.
The C-D complex increased 20\% in intensity solely due to component C,
whereas the B component decreased by an equivalent amount, simulating
constant intensity, within amplitude calibration errors, for the source as
a whole.\\
The most striking difference between the two models shown in Fig. \ref
{fig2} is the component south-east of the central component detected in
the 1982.9 data, but not required in the model for the 1997.8 data. This
component, labeled G in Fig. \ref{fig2} (right), seems to have faded out
to a level below the threshold adopted for our models. However, in the
clean-component image from the 1997.8 data (Fig. \ref{fig1}, left) there
is faint extended structure seen at the position corresponding to
component G.

\subsection{Self-similar evolution}

It is of interest to check whether the ongoing source evolution --
detected as an increase in component separation -- follows the
self-similar evolution model, which has been proposed for young radio
sources, such as GPS and CSS sources (Snellen et al. \cite{Snellen1},
\cite {Snellen2}; Snellen \& Schilizzi \cite{Snellen3},
\cite{Snellen4}).\\
Self-similar evolution of a simple two-component compact symmetric source
requires a proportional increase of the component sizes as the source
components separate from each other. An increase in the size of the
components in the image plane must be accompanied in the uv-plane by a
proportional decrease of the FWHM of the upper envelope of the amplitudes
which convolves the amplitude variation due to the beating of the two
components. These decaying fringes can be seen in Fig. \ref{fig4}, where
the upper envelope for the 1997.8 data is traced as a dotted line.
However, self-similar growth is not observed when the uv-data from epochs
1982.9 and 1997.8 are compared -- instead, we observe a decrease of the
component size as the source expands, contrary to that expected in the
self-similar growth model. This can be seen from Fig \ref{fig4} (bottom)
where the gaussian-like upper envelope from the 1997.8 fringes lies above
the 1982.9 epoch maxima, indicating that source components have shrunk in
size.  The observed shrinkage of 20\% is not due to high spatial frequency
information from self-calibrating the complete 5-GHz VSOP uv data set
before modelfitting. For the purpose of detecting changes in source
structure we flagged the HALCA baselines before performing
self-calibration.

\section{Discussion}

\subsection{Morphological classification \& age}

Compact radio sources can roughly be classified into two morphological
groups whose different appearances are believed to arise from orientation
effects. For sources showing \emph{symmetric double} structure, the radio
axis, along which the individual components are aligned, lies near the
plane of sky, whereas in sources with \emph{core-jet} morphology the radio
axis is pointing more towards the observer. In the latter case the
observed structure is highly affected by projection and relativistic
effects. In which category does 2021+614 belong?\\
The high resolution VSOP image (Fig. \ref{fig1}, left) reveals many
details not seen in earlier observations. It shows that at the higher
resolution provided by the space baselines, a compact component is visible
between component B and D at the end of a low brightness jet in the
direction of component D. Its central position relative to the two most
prominent components, B (NE hotspot/lobe) and D (SW hotspot) and the low
surface-brightness linear feature (jet) connecting it with component C (SW
lobe) suggest its identification as the central engine of activity -- the
core.\\
The presence of component A and the extended emission seen at 5~GHz
south-east of the central component are possibly not consistent with a
classification as a CSO. While component A could be outward moving plasma
emitted from the core at an earlier epoch than component B and D, it could
be argued that D is the core component, since it is the most compact
component and has the most inverted spectrum between 5 and 15 GHz and thus
the highest turnover frequency.  In addition component D is situated at
the end of a linear arrangement of source components C, E, B and A which
could be interpreted as regions of high emissivity (knots) along the path
of an outward flowing jet.
We note that Conway et al. (\cite{Conway}) give component G (Fig.
\ref{fig2}, right) and the extended emission detected east of component D
as a possible counterjet identification. They argue that the misalignment
could be explained by projection effects.  On the 5 GHz VSOP map, however,
this component is resolved out, which is not consistent with identifying
it as a compact counterjet/hotspot.\\
Our rate of separation of components B and D of $14.9 \pm 0.2$ $\mu$as/yr
obtained from a linear regression fit, corresponds to an apparent speed of
separation of $0.14\,h^{-1}\,c$ in the source reference frame. This means
that the two components were ejected $380 \pm 7$ yr ago assuming constant
velocity. With respect to the weak central component at the nucleus the
speed of separation is $\sim0.07\,h^{-1}\,c$.\\
The ``Interference Pattern Method'' of parametrizing the structural change
in the source (see Appendix A) shows that the percentage increase in the
separation along the position angle defined by the u and v stretching
factors $s$ is 3.0\%.  This increase over a timerange $\Delta t$ of 14.9
years between epoch 1982.9 and 1997.8 implies that components B and D were
ejected $t= s^{-1}\,\Delta t = 500 \pm 10$ yr ago, assuming constant
velocity. The apparent inconsistency between this value and the $380 \pm
7$ yr source age deduced above is caused by underestimation of the
uncertainty by the linear regression fit process. A conservative estimate
for the source age and its error given by the average of the two age
values, is $440 \pm 80$ yr. The corresponding separation rate and hotspot
advance speed are $ = 13 \pm 3 \, \mu$as/yr and $0.06 \pm
0.01\,h^{-1}\,c$, respectively.\\
The subluminal character of the separation speed argues in favour of a CSO
classification for 2021+614. And so do the low, total linear polarization
of the source and the absence of compact components south-west of
component D, the nucleus in the core-jet scenario.  An undetectable
counter-jet would require strong relativistic beaming effects, which would
imply high apparent expansion speeds, on the order of speed of light or
higher.
Alternatively, the source 2021+614 might be a member of the blazar group
seen face-on and observed at a extremely small angle $\theta \ll 1/\gamma$
from the line of sight, where $\gamma$ is the Lorentz factor -- resulting
in low apparent pattern speeds. However, it seems very unlikely that
2021+614 is a blazar for a number of reasons, including, the large radio
luminosity of the source together with its optical identification as an
elliptical galaxy, the stellar component in the optical spectrum and the
upper limit of 0.09 on the line flux ratio H$\beta$($\lambda_0
4865$)/[OIII]($\lambda_0 5007$) found by Bartel et al. (\cite{Bartel2}).
All these points argue convincingly for a classification as a radio
galaxy. In general, the optical properties -- spectral and morphological
-- of 2021+614 are similar to those of other (radio-loud) NLRG or
(radio-quiet) Seyfert~2 galaxies. These objects are seen edge-on,
following the Unified Models for AGN.\\
Therefore, regarding the measured increase in separation as a real
expansion, we are confident in assigning 2021+614 to the class of compact
symmetric objects.\\

\subsection{Self-similar evolution \& hotspot advance speeds}

Measurements of the hotspot advance speeds and infered kinematic ages for
CSOs trace out an interesting evolution scenario for young radio sources.  
Hotspot advance speeds for CSOs span a range from $0.06\,h^{-1}\,c$ to
$2.0\,h^{-1}\,c$ (Owsianik et al. \cite{Owsianik2}, Owsianik \& Conway
\cite{Owsianik1}, Owsianik et al. \cite{Owsianik3}, Polatidis
\cite{Polatidis}). This indicates that radio galaxies spend a few thousand
years in the GPS/CSO evolution stage. During this stage radio galaxies
apparently do not grow following a \emph{simple} self similar evolution
scheme.  This statement is based on the differences in structure between
the 1982.9 and 1997.8 models detected in 2021+614. However, the observed
decrease of the component sizes in this particular source does not reduce
the importance of the self-similar evolution model for young radio
sources. The timescales characterizing the GPS phenomenon as a whole are
measured in thousands of years.  Changing local environmental conditions
on sub-parsec scales in the NLR (Narrow Line Region) medium caused by high
density clouds could produce shock fronts which increase the compression
of the ram-pressure confined and shock-ionized radio emitting plasma in
the hotspots during short time scales of tens of years. On longer time
scales, self-similar growth is recovered because it is controlled by the
average external density of the NLR into which the radio galaxy expands
and by the power with which the jet is driven forward.\\
Calculations carried out by Owsianik et al.  (Owsianik et al.
\cite{Owsianik1}, \cite{Owsianik2} and \cite{Owsianik3}) for CSOs together
with age estimates for those sources indicate similar environmental
conditions for all of them.  Moreover, all objects studied so far are
members of the group of bright and therefore powerful GPS radio galaxies
with an intrinsic radio power output of $10^{26.0\mathrm{-}26.6}$ W/Hz at
5 GHz. Central engines powering GPS radio galaxies of similar radio
luminosities create hotspots with similar internal pressure under similar
environmental conditions (external density, density profile, magnetic
field). In particular, the internal pressure and size of the working
surface of the hotspot determine the kinetic power transported by the jet,
which is responsible for generating the observed hotspot advance speeds.
Therefore, not surprisingly, all hotspot advance speeds are of the same
order.\\

\section{Conclusions}

We provide strong evidence that 2021+614 is one of a small group of
compact symmetric sources for which speeds of separation have been
measured.  All have apparent ages of a few hundred to a few thousand
years, indicating their youthfulness.\\
In the case of 2021+614 we measure a separation of $16.1\,h^{-1}$ pc and a
separation rate of $0.12 \pm 0.02\,h^{-1}\,c$ between the two dominant
components.  These components are associated with lobes and/or hotspots.
The hotspot-advance speed is $0.06 \pm 0.01\,h^{-1}\,c$. From separation
and separation rate measurements we deduce an apparent age of $440 \pm 80$
yr. All results are measured in the source reference frame.\\
We do not observe self-similar growth in 2021+614 over a timerange of 15
yr but we argue that this does not rule out the self similar evolution
scheme for young radio sources over longer timescales.

\appendix \section{Errors in separation measurements}

In order to generate an error estimate for the separation rate of 14.9
$\mu$as/yr deduced from 5 GHz data shown in Fig. \ref{fig3} we need to
consider the uncertainties for each individual data point.  This is not a
simple issue since we do not have the original data from which the two
data points at 1982.9 and 1987.7 were deduced.\\
However, based on the assumption that a linear relation fits the
separation measurements well and errors for individual data points are
equal, a linear regression fit can provide error estimates for the
separation measurement. Adopting this procedure we obtain an uncertainty
of 2.1 $\mu$as for separation measurements and 0.2 $\mu$as/yr for the
separation rate at 5 GHz.  This method, however, excludes the possibility
of an independent estimate for the goodness-of-fit. \\
Statistically correct treatments, such as elliptical gaussian fits in the
image plane (e.g. AIPS task \verb+JMFIT+) yield very small positional
uncertainties for high dynamic range images (Fomalont \cite{Fomalont2}).  
The image in Fig. \ref{fig2} (middle) allows the separation between
component B and D to be determined with an accuracy an order of magnitude
better as compared to the linear regression fit.\\
In order to circumvent this inconsistency between error estimations for
the separation measurements and to obtain a second, independent source age
estimate we developed a new method which we present below.

\subsection{The ``Interference Pattern Method''}

\begin{figure}
\hbox{\epsfxsize=8.7cm \epsfbox{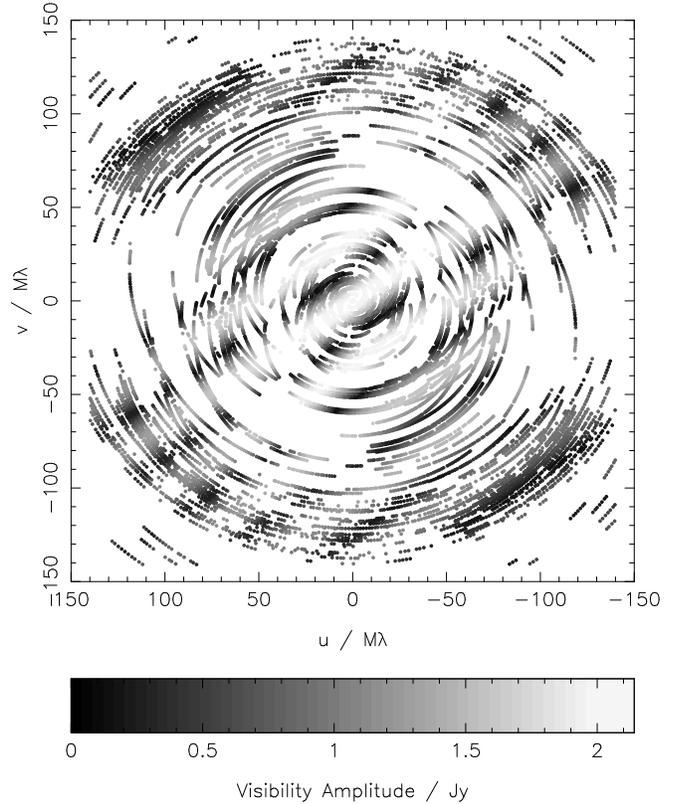}}
\caption{The uv-data sampled during the 1997.8 observing run gridded on
the $561 \times 561$ pixel mesh. These data include only components B, D
and the noise left after having subtracted components A, B', E, F and C.
The characteristic interference pattern shows alternating maxima and
minima along position angle 33.5$\degr$ and represents the two-dimensional
reality behind the one-dimensional projection plots shown in Fig.
\ref{fig4}.}
\label{fig5} 
\end{figure}

Our goal is to determine changes in source structure occured between
1982.9 and 1997.8 directly in the uv plane, using the amplitude
interference patterns. These changes are parametrized by a two-dimensional
stretching factor whose components along the u and v-axis are $s_{\mathrm
u}$ and $s_{\mathrm v}$, and by an amplitude correction factor $s_{\mathrm
a}$. Our method provides estimates for these quantities and for their
uncertainties. In this way, interpreting the observed shift in the
position of the visibilty amplitude minima between the two epochs -- best
seen in Fig. \ref{fig4} -- as real expansion, we are able to determine the
source age, the separation rate and the hotspot advance speed.\\

\subsection{Parametrizing the time evolution/structural change}

Conway's 1982.9 data set was obtained during a 9 hour global VLBI
observing run using 4 antennas in the USA and the 100-m dish in
Effelsberg. Consequently, compared to our 1997.8 data set the uv coverage
is sampled more sparsely and the two observations measure different
spatial frequencies.\\
In order to obtain two data sets suitable for detecting an increase in
separation between the two most prominent components we prepared the two
uv data sets as follows: \emph{1982.9} -- a simple two-component model,
comprising components B and D from Conway's six-component model for the
1982.9 observations, was Fourier-transformed back into the uv-plane. The
model visibility amplitudes were calculated on a $561 \times 561$ pixel
grid with mesh size of 0.5 M$\lambda$ ranging from $-140.0$ to $140.0$
M$\lambda$ in both the u and v-dimension. This grid spacing was fine
enough to detect a decrease in the position of the first minimum of a few
percent; \emph{1997.8} -- all the model components except B and D were
uv-subtracted from the 1997.8 observations.  In order to take account of
the flux density variability of the components, the relative intensities
of B and D were adjusted to match Conway et al's values. Then, the
visibility amplitudes for the 1997.8 data, sampled along the uv-tracks
shown in Fig. \ref{fig2} (left), were gridded onto the $561 \times 561$
pixel mesh.\\
The gridding of the visibility data was done by assigning the visibility
value to the nearest grid point. Multiple assignments near the uv-plane
centre were averaged together, but no special weighting was applied to
those data points. Fig. \ref{fig5} shows a grey-scale plot of visibilty
amplitudes measured during the 1997.8 observation, prepared as explained
above and gridded onto the $561 \times 561$ pixel mesh.\\
After having prepared the data sets we assumed that all remaining
differences between 1997.8 ``data'' and 1987.9 ``model'' are either due to
real expansion or a residual multiplicative amplitude correction factor,
$s_\mathrm{a}$. The parametrization of the change in separation was done
by applying multiplicative factors, $s_\mathrm{u}$ and $ s_\mathrm{v}$, to
the sampled uv-points and associating the amplitude value of the original
point with the new coordinates. Such an evolution model is, strictly
speaking correct, only for unresolved source components which separate
from or contract towards each other. For a pair of resolved components,
modelled by elliptical gaussians, the resulting fringe pattern is
convolved with an elliptical gaussian whose major and minor axis FWHM are
inversely proportional to the corresponding parameters in the image plane.  
We investigated the impact of this simplification on the fitting process
and concluded that it is minor and therefore negligible at the levels of
accuracy involved.

\subsection{Fitting for the scale factors $s_\mathrm{u}$, 
$s_\mathrm{v}$ and $s_\mathrm{a}$}

The $\chi^2$ parameter measuring the goodness of fit was defined as the
sum of the squared differences of the visibility amplitudes:
\begin{displaymath} 
\chi^2 = \sum_{i=1}^N \frac
{[s_\mathrm{a}\,A_1(s_\mathrm{u}\,u_i, s_\mathrm{v} v_i)-
A_2(u_i,v_i)]^2}{\sigma_i^2}, 
\end{displaymath} 
where the indices 1 and 2 refer to the 1982.9 model and 1997.8 data,
respectively; $N$ is the total number of gridded visibility measurements;
$\sigma_i$ is the error associated with the visibilty amplitude data
point. Values $> 1$ for $s_\mathrm{u}$ and $s_\mathrm{v}$ indicate inward
shifting minima as the source evolves. The result provided by the fitting
routine gave a value of a few tens for the minimum reduced $\chi^2$ rather
than unity. This fit is too poor to use the $(\chi^2+1)$-contour
projections onto the parameter axes to determine the 1-$\sigma$
uncertainties for $s_a$, $s_\mathrm{u}$ and $s_\mathrm{v}$. Nevertheless,
the procedure outlined above provided us the best-fit values for the
overall amplitude correction factor, $s_\mathrm{a} = 0.88$ and for the
scale factors in u and v-direction, $s_\mathrm{u} = 1.0445$ and
$s_\mathrm{v} = 1.0202$.\\
The percentage increase in separation $s$ along the position angle defined
by the u and v stretching factors $s_u$ and $s_v$ is given by
\begin{displaymath}
s=[(s_\mathrm{u}-1)^2\sin^2\psi_1+(s_\mathrm{v}-1)^2\cos^2\psi_1]^{1/2}
\end{displaymath} and is 3.0\%, where $\psi_1$ is the angular polar
coordinate of component B with respect to D at epoch 1987.9.\\
The most difficult problem to tackle during the fitting process is to
ensure that differences in component size and shape between different
observations do not influence substantially the best-fit values of the
parameters. The relatively high residual correction of 0.88 needed for the
amplitude parameter, and the difference between $s_\mathrm{u}$ and
$s_\mathrm{v}$, are due to different component sizes at the two epochs. A
more sophisticated method would take these details into account and be
more widely applicable. However, for our purpose, which is limited to the
determination whether or not structural change in 2021+614 can be
explained by subluminal motion of components, the systematic errors
introduced are not relevant.

\subsection{Bootstrap method for estimating errors in the scale factors
$s_\mathrm{u}$ and $s_\mathrm{v}$.}

The bootstrap method is a very powerful error estimation technique
applicable when there is not enough knowledge about the nature of the
measurement errors to do a proper Monte Carlo simulation. We use this
method to estimate the errors for the parameters $s_\mathrm{u}$ and
$s_\mathrm{v}$.\\
For the 1997.8 observation we generated one hundred uv-data samples each
containing a different 50\% of the original data. The visibility
measurements in each sample were selected randomly. We fitted the one
hundred data sets to Conway's model visibilites as outlined in the
previous section for the whole data set, but this time fitting only the
two scale factors $s_\mathrm{u}$ and $s_\mathrm{v}$, and setting
$s_\mathrm{a}$ to its best-fit value of 0.88. The standard deviations of
the best-fit parameters provide the errors associated with these
quantities. We found $s_\mathrm{u} = 1.0443 \pm 0.0005$ and $s_\mathrm{v}
= 1.0212 \pm 0.0010$. This means that using 50\% of the available data we
are able to determine a 4\% decrease in separation of the minima along the
u-axis and a 2\% decrease along the v-axis at an accuracy level of 5\% and
10\%, respectively -- \emph{clear evidence for real motion}.  The
difference in accuracy is due to the higher percentage increase along the
u-axis.  The values found agree with the scale factors determined using
all the data, reported at the end of the previous section. The errors
improve by a factor of $1/\sqrt{2}$ if twice as many amplitude
measurements are used.\\
In order to check the results we obtained for the scale-factor errors
$s_\mathrm{u}$ and $s_\mathrm{v}$ we can carry out a simple calculation.
3\% is the minimum change in scale factor which moves the uv-points by at
least one pixel for uv-loci with $\sqrt{u^2 + v^2}> 15$ M$\lambda$.  With
about 5200 uv-points obeying this constraint, and taking into account that
each measured and gridded point contributes significantly to the shift
determination, the accuracy of the shift determination is improved by a
factor of $1/\sqrt{5200}$, i.e. $0.03/\sqrt{5200} \simeq 0.0004$, which in
first approximation agrees with the errors reported above.

\begin{acknowledgements}

This research was supported by the European Commission's TMR Programme,
``Access to Large-Scale Facilities'', under contract No. ERBFMGECT950012.  
We gratefully acknowledge the VSOP Project, led by the Japanese Institute
of Space and Astronautical Science in cooperation with many organizations
and radio telescopes around the world.  The National Radio Astronomy
Observatory is a facility of the National Science Foundation operated
under cooperative agreement by Associated Universities, Inc.  We thank
J.E. Conway for providing the component model for the 1982 observation and
the absolute separation measurements for both of his observations.  
Special thanks go to R.C. Vermeulen and K.I. Kellermann for donating
self-calibrated uv-data sets from their 2 cm observations of 2021+614. We
thank the referee Hugh D. Aller for careful reading and helpful
suggestions.

\end{acknowledgements}

\end{document}